\documentclass{cimento}

\usepackage{graphicx}  
\title{Strangeness in the nucleon: what have we learned?}
\author{A.~W.~Thomas\from{ins:A}, 
P.~E.~Shanahan\from{ins:A} \atque
R.~D.~Young\from{ins:A}}
\instlist{\inst{ins:A} ARC Centre of Excellence for Particle Physics at the Tera-scale 
and \\ CSSM, School of Chemistry and Physics,\\
University of Adelaide, Adelaide SA 5005 Australia}
\PACSes{
\PACSit{14.20.Dh} {Protons and neutrons} 
\PACSit{12.38.Gc} {Lattice QCD calculations}
\PACSit{13.40.Gp} {Electromagnetic form factors}
}

\begin{document}

\maketitle

\begin{abstract}
We review the state of our knowledge concerning the contribution of strange 
quarks to various nucleon properties. In the case of the electric and magnetic 
{}form factors, the level of agreement between theory and experiment is very 
satisfactory and gives us considerable confidence in our capacity to make 
reliable calculations within non-perturbative QCD. In view of the 
importance of the scalar form factors to the detection of 
dark matter candidates 
such as neutralinos, we place a particular emphasis on the determination of 
the $\pi N$ and strange quark sigma commutators.
\end{abstract}

\section{Introduction}
As the only serious candidate for a fundamental 
theory of the strong interaction, 
QCD presents some remarkable challenges. The quantisation of such a highly 
non-linear quantum field theory is extremely difficult and at present only 
lattice QCD has a claim of significant success. In assessing our progress 
towards a complete understanding of QCD it is critical to check that key 
calculations actually agree with experiment. 
{} For QCD the strange form factors of the nucleon 
occupy a position of comparable importance to that of the Lamb shift 
in the history of QED. 
While lattice QCD has accurately described a number of 
valence quark dominated hadronic 
properties, the strange form factors of the nucleon 
can only arise through quantum 
fluctuations, in which a strange-anti-strange pair briefly bubble into and out 
of existence. Thus the calculation of the strange form factors and 
their verification by experiment is of fundamental importance.

In the next section we briefly review the state of play with respect to the
strange electric and magnetic form factors of the nucleon, which have been 
the focus of intensive experimental effort for the past two decades.
We then turn to the 
strange sigma commutator (the strange scalar form factor, $\sigma_s$), for 
which the best
value has undergone a major shift in the last two or three years.
After summarising recent work on the determination of $\sigma_s$, 
including a parallel analysis of data on the mass of the $H$-dibaryon, 
we explain the relevance to dark matter searches.
The final section is devoted to some concluding remarks.

\section{Vector form factors}
It is now more than 20 years since it was realized that parity 
violating electron scattering (PVES) could provide a third, 
independent constraint on 
the vector form factors of the nucleon, thus allowing one to solve for the 
strange vector matrix elements~\cite{Mckeown:1989ir}. 
{}Following a series of state-of-the-art 
measurements at MIT-Bates, Mainz and JLab (for a recent review see 
Ref.~\cite{Paschke}), as well as a systematic study 
of the relevant radiative 
corrections~\cite{Blunden:2011rd,Gorchtein:2008px} 
and a careful global analysis~\cite{Young:2006jc}, we now 
know that the strange magnetic form factor is at most a few percent of the 
proton magnetic form factor at low-$Q^2$ and the strange electric radius 
is also at most a few percent of the proton charge radius.

Remarkably, it is now twenty years since the initial studies of nucleon 
vector form factors 
in lattice QCD, by Leinweber and collaborators~\cite{Leinweber:1990dv}.
In combination with chiral extrapolation 
of lattice data using finite range regularisation, 
indirect techniques developed by 
Leinweber and Thomas~\cite{Leinweber:1999nf} led to 
a very precise determination of both the strange 
magntic moment of the proton and its strange charge 
radius~\cite{Leinweber:2004tc} some 15 years later. 
Although it initially seemed as though these calculations disagreed with the 
PVES data, it is now clear that the agreement is excellent. Furthermore, 
precise, direct calculations of these form factors from 
the Kentucky group in just 
the last two years~\cite{Doi:2009sq} agree very well with the 
results of the earlier indirect work 
and with the experimental results. 

At present, the accuracy of the theoretical calculations exceeds that of the 
best experiments by almost an order of magnitude -- a remarkable exception in 
strong interaction physics. Clearly the challenge is there for a clever new 
idea to take us beyond the current experimental limitations. Nevertheless, 
this future challenge should not blind us to the tremendous achievements 
thus far and especially to the fact that QCD has passed its equivalent of 
the ``Lamb shift test'' with flying colours.

\section{Sigma commutators}
The so-called $\pi N$ sigma commutator
\begin{equation}
\sigma_{\pi N} = m_l \langle p | (\bar{u}u \, + \, \bar{d} d) | p \rangle \, 
\equiv  m_l \frac{\partial M_N}{\partial m_l} \, ,
\end{equation}
with $m_l= (m_u + m_d)/2$ and the strange sigma commutator
\begin{equation}
\sigma_s = m_s \langle p | \bar{s}s | p \rangle \, 
\equiv m_s \frac{\partial M_N}{\partial m_s} \, ,
\end{equation}
not only tell us directly how much the corresponding quark masses 
contribute to the mass of the nucleon but they also 
constitute direct measures of chiral symmetry breaking in QCD. They have 
therefore been of great interest ever since the phenomenological 
importance of chiral symmetry was realized.

One often used measure of the relative importance of the strange quarks in
nucleon structure is the ratio $y$:
\begin{equation}
y = \frac{2 \, \langle p| \bar{s} s | p \rangle}{\langle p| (\bar{u}u + \bar{d} d )
| p \rangle } = \frac{m_l}{m_s} \frac{2 \,  \sigma_s}{\sigma_{\pi N}} \, .
\end{equation}
Early calculations of quantities such as this relied on the naive 
application of formulas 
for the octet baryon masses based upon first order breaking of flavour SU(3) 
symmetry. For example, the {\it non-singlet} combination $\sigma_0$:
\begin{equation}
\sigma_0 = m_l \langle p | \bar{u} u \, + \, \bar{d}d \, -  \, 2 
 \bar{s}s | p \rangle \, ,
\end{equation}
was found from such an analysis to 
be $\sigma_0 = 36 \pm 7$ MeV~\cite{Borasoy:1996bx}. 
Then, after the $\pi N$ sigma commutator is deduced from the historical method 
of pion-nucleon dispersion relations to get the value of the scalar form factor 
at the unphysical Cheng-Dashen point ($t=2m_\pi^2$) 
and then corrected for form factor 
effects to the point $t=0$, the ratio 
$y$ may be found as $ (\sigma_{\pi N} - \sigma_0)/\sigma_{\pi N}$.

This method has a number of practical problems. First there is some controversy over 
the value at the Cheng-Dashen point extracted from dispersion relations. Second, 
the form factor correction is sizeable and not model independent. Finally, the 
numerical difference between $\sigma_0$ and $\sigma_{\pi N}$ is small 
compared with the errors in both of them. As a result, values quoted for 
$y$ have typically ranged between 0 and 0.4, with $\sigma_s$ itself being taken 
to be of order 300MeV for many years. Indeed, this large estimate, combined 
with early overestimates of the strange vector from factors played a crucial 
role in motivating the parity violation experiments discussed above.

Since the strange sigma commutator may be interpreted as the contribution 
to the mass of the nucleon from the strange quark, a value as large as 300 MeV 
would indeed be remarkable. It would mean that almost a third of 
the nucleon mass arises from 
quarks that are non-valence. This appears incompatible with the widely used 
constituent quark models, for example. It is not surprising that this has 
motivated an enormous amount of theoretical interest in pinning down the 
empirical value of $\sigma_s$ as accurately as possible.
However, it is in the search for dark matter that
the value of $\sigma_s$ has had its most immediate
impact~\cite{Giedt:2009mr}.

\subsection{Importance in the search for dark matter}
In the minimal supersymmetric extension of the Standard Model, constrained by
all particle physics and WMAP data, the so-called CMSSM, the favoured
candidate for dark matter is the neutralino~\cite{Jungman:1995df},
a weakly interacting fermion
with mass of order of a hundred GeV or more. For the old values of
$\sigma_s \sim 300$MeV, its dominant interaction 
with the nucleon was through the strange
quark. The old method of determining $\sigma_s$, through the difference between 
$\sigma_{\pi N}$ and $\sigma_0$, led Ellis, Olive and collaborators to 
call desperately for
a more accurate determination of $\sigma_{\pi N}$, not for its own sake 
but in order to pin down
$\sigma_s$~\cite{Ellis:2008hf}.

We return to this topic below, after reviewing more modern methods to 
determine $\sigma_s$, based upon lattice QCD and in particular 
the careful chiral analysis 
of lattice data.

\subsection{Lattice QCD}
Over the past 15 years there have been 
extensive studies of the sigma commutators 
within lattice QCD, with the initial calculations tending to support the 
large values for $\sigma_s$. On the other hand, 
in the absence of 
sufficient computing power 
to directly calculate hadron properties at the physical quark masses, 
we have been fortunate to be given data that has opened new methods 
to determine $\sigma_s$. In particular, a wealth of data has 
been generated for a variety 
of hadronic properties {\it within QCD} as 
a function of the masses of the quarks. 
This is of course information that Nature 
cannot give us but which is nevertheless 
an invaluable guide to how QCD actually 
works~\cite{Detmold:2001hq}. From those studies it is clear that for 
whatever reason (and we suggest one 
below) the properties of baryons and non-Goldstone mesons made of light quarks 
behave exactly as one would expect in a constituent 
quark model once the pion mass 
is above about 0.4 GeV. That is, all the famous, rapid, non-analytic variation 
associated with Goldstone boson loops is seen to 
disappear in this region~\cite{Leinweber:1998ej,Leinweber:1999ig}.

There has been some speculation that this 
scale (which corresponds to a quark mass 
around 40 MeV) may have something to do with 
the size of an instanton. However, it 
is clear that such behaviour does emerge naturally 
if one takes into account the 
finite size of the hadrons, which necessarily suppresses meson loops at large 
momentum transfer and at large mass for the Goldstone bosons. Put simply, 
meson loops are suppressed when the corresponding 
Compton wavelength is smaller than 
the size of the hadron emitting or absorbing the meson. 
Given that a typical hadron 
size is 1fm, and the Compton wavelength of a meson of 
mass 0.4 GeV only 0.5fm, one 
has a natural explanation.

This qualitative lesson from QCD itself, leads us to anticipate 
that the contribution 
of strange quarks to nucleon properties should be suppressed, 
because the mass of the kaon 
is 0.5 GeV  -- i.e., it is above the critical scale. 

This idea has also been developed into a method of analysis which 
has enabled  quantitative advances in the 
calculation of hadron properties using finite range regularization 
(FRR)~\cite{Young:2002ib}. This 
technique allows one to effectively resum the 
chiral expansion of hadronic properties 
and to accurately describe the variation of properties such as the mass of the 
baryon octet over a much larger range of quark masses 
than expected within naive 
chiral perturbation theory. 

\subsection{Formal expansion of the sigma commutator}
The literature is dominated with suggestions that the sigma commutators 
are dominated by the leading terms in the chiral expansion of the 
nucleon mass. For example, the chiral coefficient $c_1$, which yields 
the term proportional to $m_\pi^2$ in the standard expansion of 
$m_N$, is often refered to as the term that gives $\sigma_{\pi N}$.
Of course, it has been known for decades that the leading non-analytic 
(LNA) term, proportional to $m_\pi^3$ is almost as big 
(and opposite in sign) but the discussion 
usually stops there. While higher order terms in the chiral expansion 
are not model independent, within FRR with the formulae fit to lattice 
data for $m_N$, it has been shown that the value of $\sigma_{\pi N}$ 
extracted is independent of the model used for the 
regulator~\cite{Leinweber:2003dg,Leinweber:2000sa}. It therefore seems 
reasonable to use FRR chiral perturbation theory to serve as a guide 
to the importance of higher order terms in the expansion of 
$\sigma_{\pi N}$ in powers of $m_\pi$.

As an example, using a dipole form factor one finds for the expansion 
of the $\pi N$ loop which yields the leading non-analytic (LNA) behaviour 
of $m_N$:
\begin{equation}
\delta M_N = c_{{\rm LNA}} \left( \frac{\Lambda^3}{16} - \frac{5 \Lambda}{16}
m_\pi^2 + m_\pi^3 - \frac{35}{16 \Lambda} m_\pi^4 +\ldots \right) \, ,
\end{equation}
where $c_{{\rm LNA}} = -3 g_A^2/32 \pi f_\pi^2$ and $\Lambda$ is the 
dipole mass parameter. From this one may easily show that the corresponding 
contribution to $\sigma_{\pi N}$, at the physical pion mass, is
\begin{equation}
\delta \sigma_{\pi N} = 
35 \Lambda \, -23 \, + \frac{9.6}{\Lambda} \, - \frac{3}{\Lambda^2} + \ldots
\end{equation}
where $\Lambda$ is in GeV and the result in MeV.
Thus, for a typical value of $\Lambda \sim 1$GeV, we see that the $m_\pi^4$ 
term contributes almost 10 MeV and so, even for $\sigma_{\pi N}$ the series 
expansion is very slowly convergent.

However, for $\sigma_s$, where the LNA  behaviour in $m_K$ has the same 
form, with a very similar value of $c_{{\rm LNA}}$, the much larger 
mass ($m_K/m_\pi \sim 3.5$ ) means that the cubic term is of order -1 GeV 
and the quartic term around +1.4 GeV. Clearly, the expansion 
of $\sigma_s$ in powers of $m_s$ about zero is badly divergent and therefore 
useless. On the other hand, using a FRR to resum the series in a fit to 
lattice data for $m_N$ versus $m_\pi$ over a large range does lead to 
results that are independent of the model used. This is a powerful 
technique.

\subsection{Modern values of the sigma commutators}
Once one has an accurate parametrization of the 
mass of the nucleon as a function 
of pion and kaon mass, based on a fit to modern lattice data for the nucleon 
octet using FRR, it is trivial to extract the light quark sigma commutators 
by differentiation, using the Feynman-Hellmann theorem. It is this approach 
which has recently shown that $\sigma_s$ is almost an order of magnitude 
smaller than had been generally believed for 20 years~\cite{Young:2009zb}. 
Indeed, in a systematic study of the masses of the octet baryons, using 
chiral perturbation theory with FRR, Young and Thomas found the value 
$\sigma_s = 31 \pm 15 \pm 4 \pm 2$MeV~\cite{Young:2009zb}. Combining 
this with the value they 
obtained for $\sigma_{\pi N} = 47 \pm 8 \pm 1 \pm 3$MeV, 
this yields a value for $y \sim 0.05$.

It is interesting to note that a small value for $\sigma_s$ 
had been published by 
the Adelaide group some years earlier, 
in the context of the search for variations in 
fundamental constants~\cite{Flambaum:2004tm}. It is well known that 
in many grand unified theories a change in $\alpha$ with time leads 
to a much larger change with time of quark masses. That this may 
have observable consequences in modern precision measurements of 
atomic spectroscopy means that one needs to know the variation of 
parameters such as the nucleon mass and magnetic moment with both 
light and strange quark masses. Based solely on the contribution 
from $\pi, K$ and $\eta$ loops, Flambaum {\it et al.} reported a 
value of $\sigma_s \sim 10$MeV. The value reported above, in 
which the meson loops are calibrated by fitting accurate octet mass 
data, represents a natural improvement in the earlier estimate. However, 
the fact that $\sigma_s$ is {\it much} smaller than had been believed 
is a feature of both calculations.

In direct contrast with the early lattice simulations, which tended 
to reinforce the large values of $\sigma_s$, 
very similar conclusions have been reached by several other modern lattice 
simulations~\cite{Ohki:2008ff,Toussaint:2009pz,Ohki:2009mt}, 
with most agreeing that the value is between 20 and 50 MeV. 
Although at first sight it is shocking that there can be such a large shift 
in a fundamental property of the nucleon, it is quite a 
common phenomenon when it comes to  
fundamental parameters that apparent convergence can be followed by a shift 
by far more than the quoted uncertainties when a new technique becomes 
available.

Recent work from UKQCD supports earlier suggestions 
that the underlying reason for 
erroneous large values of $\sigma_s$ in the early lattice studies is operator 
mixing. Indeed, after a careful analysis of the operator mixing under 
renormalization, Bali {\it et al.} recently reported 
$\sigma_s = 11 \pm 13 \, \stackrel{+9}{-3}$MeV~\cite{Bali:2011ks}. 
We note that 
this was at a light quark mass somewhat larger than the physical value and 
correcting for this should raise the final value by several MeV.

The European Twisted Mass Collaboration also reported a new result for 
$y$ recently~\cite{Dinter:2011zz}, 
namely $y = 0.066 \pm 0.011 \pm 0.002$.  
This agrees very well with 
the determination of Young and Thomas quoted above.

\subsection{$H$-dibaryon mass}
In the context of recent new lattice results which suggested that the 
$H$-dibaryon might indeed be bound with respect to the $\Lambda-\Lambda$ 
threshold~\cite{Beane:2010hg,Inoue:2010es}, Shanahan {\it et al.} recently 
built on the work of Mulders and Thomas~\cite{Mulders:1982da} to 
make a detailed FRR analysis 
of both the $H$ data and the full octet baryon 
data-set~\cite{Shanahan:2011su}. Their result, namely that at the physical 
quark masses the $H$ is most likely slightly unbound (by $13 \pm 14$MeV), 
is of great interest both because of its potential implications for 
the equation of state of dense matter and in connection with the 
enhancement above $\Lambda-\Lambda$ threshold already reported 
experimentally~\cite{Yoon:2007aq}. It is clearly of great importance 
that the latter be pursued in experiments at the new J-PARC facility.

In the present context that study is of direct interest because the  
analysis of the full octet data-set actually 
led to a much improved error band 
on the values of $\sigma_{\pi N}$ and $\sigma_s$. For details of the 
systematic study of potential sources of error in the analysis of 
Shanahan {\it et al.} we refer to Ref.~\cite{Thomas:2011cg}. Suffice 
it to say that under variations of the various chiral parameters 
and the form of the UV regulator, the 
values of $\sigma_{\pi N}$ varied between 42 and 51 MeV, with a 
statistical 
error of order 5 MeV. In fact, the only major shift was for the sharp cut-off 
which is somewhat unphysical. Omitting that and combining the errors 
in quadrature the result is $\sigma_{\pi N} = 44 \pm 5$MeV. For 
$\sigma_s$ the variation with regulator was negligible and the value 
deduced was $21 \pm 7$MeV. This is arguably the most reliable current 
determination. For these values of $\sigma_{\pi N}$ and 
$\sigma_s$ and using Eq.~(3)   
we find $y = 0.04 \pm 0.02$.

\subsection{Implications for dark matter searches}
It is in the search for dark matter that 
the new value of $\sigma_s$ has had its most immediate 
impact~\cite{Giedt:2009mr}. In the context of the CMSSM, with the 
neutralino as the candidate for dark matter, the large values of 
$\sigma_s$ meant that strange quarks dominated the dark matter 
cross section.
On the other hand, the new value reported above changes the situation 
dramatically. No longer do the strange quarks dominate the 
cross section. Not only is the expected cross section reduced 
by an order of magnitude compared with earlier optimistic expectations but 
the values found are far more accurate. We refer to the work of 
Giedt {\it et al.} for more details~\cite{Giedt:2009mr}.

\section{Conclusion}
We have seen that the role played by strange quarks in the structure of 
the nucleon is important for a multitude of reasons. In any analysis 
of the electromagnetic form factors of the nucleon it is critical to 
know the strange quark contribution. After two decades work it is now 
clear that this contribution is rather small but most important that it 
is in excellent agreement with the values given by QCD.

In contrast with the suggestion that strange quarks might contribute as 
much as a third of the mass of the nucleon, sophisticated modern 
calculations based on lattice QCD and FRR chiral perturbation theory 
have shown that the contribution is nearer to a few percent. This is a 
remarkable shift in such a fundamental quantity and it has profound 
consequences for the search for dark matter as well as for possible 
changes in fundamental ``constants'' of Nature.

We close with a brief remark on a serious conundrum. The classic analysis of 
the nucleon mass by Shifman {\it et al.}~\cite{Shifman:1978zn} suggests 
that each heavy flavour contributes of the order of 70 MeV to the 
mass of the nucleon. In contrast, as we have shown, the three active flavours 
($u,d$ and $s$ ) each contribute around 20--30 MeV to 
the mass of the nucleon. Understanding 
how it is possible that the contributions from $c,b$ and $t$ quarks 
could be two or three times larger, if this is indeed correct, would be a 
fundamental contribution to our knowledge of QCD and hadron structure.

\acknowledgments
This work was supported by the University of Adelaide and by the
Australian Research Council through an 
Australian Laureate Fellowship (AWT: FL0992247), through
an ARC Discovery Grant, DP110101265 (RDY),
and by the ARC Centre of Excellence for Particle Physics at the Tera-scale.

\end{document}